\documentclass[aps,preprint,showpacs]{revtex4}
\usepackage{graphicx}

\begin{document}

\title[Accelerating Universe in a Big Bounce Model]{Accelerating Universe in a Big Bounce Model}
\author{Beili Wang, Hongya Liu\footnote{Corresponding author, hyliu@dlut.edu.cn}, Lixin Xu}
\affiliation{Department of Physics, Dalian University of Technology \\ 
Dalian, Liaoning, 116024, P. R. China} 


\begin{abstract}
Recent observations of Type Ia supernovae provide evidence for the
acceleration of our universe, which leads to the possibility that the
universe is entering an inflationary epoch. We simulate it under a ``big
bounce'' model, which contains a time variable cosmological ``constant''
that is derived from a higher dimension and manifests itself in 4D spacetime
as dark energy. By properly choosing the two arbitrary functions contained
in the model, we obtain a simple exact solution in which the evolution of
the universe is divided into several stages. Before the big bounce, the
universe contracts from a $\Lambda $-dominated vacuum, and after the bounce,
the universe expands. In the early time after the bounce, the expansion of
the universe is decelerating. In the late time after the bounce, dark energy
(i.e., the variable cosmological ``constant'') overtakes dark matter and
baryons, and the expansion enters an accelerating stage. When time tends to
infinity, the contribution of dark energy tends to two third of the total
energy density of the universe, qualitatively in agreement with observations.
\end{abstract}

\maketitle
\section{Introduction}

Recently, more and more observations of Type Ia supernovae suggest that the
observable universe is presently undergoing an accelerating expansion,$%
^{1,2} $ which is contrary to what has always been assumed that the
expansion is slowing down due to gravity with a positive decelerating
parameter $q$. This, together with the recent high precision measurement of
the cosmic microwave background (CMB) fluctuations,$^{3,4}$ indicates the
existence of dark energy characterized by a negative pressure, contributing
with about $70\%^{1,2,5,6}$ of the totel energy density of the universe (the
others being essentially dark matter and baryons). The simplest model for
dark energy is the cosmological constant $\Lambda $. However, because of the
so-called cosmological constant problem,$^{7,8}$ one wishes to have a
dynamically decaying $\Lambda $ or an evolving large-scale scalar field
known as quintessence$^{8,9}$ to explain the acceleration of our universe.

It is of great interest that our conventional universe is embedded in a
higher-dimensional world as required in the Kaluza-Klein theories and the
brane world theories. In this paper, we consider a five-dimensional
cosmological model presented by Liu and Wesson.$^{10}$ Rather than the ``big bang'' 
singularity of the standard cosmology, this 5D model is characterized by a ``big bounce'', at
which the ``size'' of the universe is finite. Before the bounce the universe
contracts and after the bounce the universe expands. This model is 5D
Ricci-flat, implying that it is empty viewed from 5D. However, as is known
from the induced matter theory,$^{11,12}$ 4D Einstein equations with matter
could be recovered from 5D Kaluza-Klein equations in apparent vacuum. This
approach is guaranteed by Campbell's theorem that any solution of the
Einstein equations in N- dimensions can be locally embedded in a Ricci-flat
manifold of (N+1)- dimensions.$^{13}$ An important result of the 5D bounce
model is that a time variable cosmological ``constant'' can be isolated out,
in a natural way, from the induced 4D energy-momentum tensor.

The 5D bounce solution contains two arbitrary functions $\mu (t)$ and $\nu
(t)$. It was shown in Ref. [10] that by properly choosing $\mu (t)$ and $\nu
(t)$, one can obtain exact solutions suitable to describe both the
radiation-dominated universe and the matter-dominated universe as in the
standard FRW models. In this paper we will show that by properly choosing $%
\mu (t)$ and $\nu (t)$, we can also obtain exact solutions suitable to
describe our present accelerating universe which is dominated by dark energy.

\section{An accelerating universe model}

The 5D cosmological solution reads$^{10}$

\begin{eqnarray}
dS^{2} &=&B^{2}dt^{2}-A^{2}\left( \frac{dr^{2}}{1-kr^{2}}+r^{2}d\Omega
^{2}\right) -dy^{2},  \nonumber \\
A^{2} &=&(\mu ^{2}+k)y^{2}+2\nu y+\frac{\nu ^{2}+K}{\mu ^{2}+k}, 
\nonumber \\
B &=&\frac{1}{\mu }\frac{\partial A}{\partial t}\equiv \frac{\dot{A}}{\mu }.
\label{5Dmetric}
\end{eqnarray}%
Here $\mu =\mu (t)$ and $\nu =\nu (t)$ are arbitrary functions, $k$ is
the three-dimensional curvature index $(k=\pm 1,0)$, and $K$ is a constant.
This solution satisfies the 5D equations $R_{AB}=0$ $(A,B=0123;5)$. So one
has $R=0$ and $R^{AB}R_{AB}=0$. The 5D Kretschmann invariant is found to be 
\begin{equation}
I=R_{ABCD}R^{ABCD}=\frac{72K^{2}}{A^{8}}.  \label{5DKretschmann}
\end{equation}%
which shows that K determines the curvature of the five-dimensional
manifold. This solution was firstly derived by Liu and Mashhoon in a
different notation in Ref. [14]. The bounce property was firstly discussed by
Liu and Wesson.$^{10}$ Further studies concerning the bounce singularity and
other properties can be found in Ref. [15-20]. This solution can also be
used to construct exact brane cosmological models.$^{21}$

The 4D part of the empty 5D metric (\ref{5Dmetric}) gives an exact solution
of the 4D Einstein equations with an effective or induced energy-momentum
tensor. It was shown in Ref. [10] that this energy-momentum tensor could be
modeled by a perfect fluid with density $\rho $ and pressure $p$, plus a
cosmological term $\Lambda $:

\begin{equation}
^{(4)}T_{\alpha \beta }=(\rho +p)u_{\alpha }u_{\beta }+(\Lambda -p)g_{\alpha
\beta }.  \label{Einsteintensor:}
\end{equation}
where $u^{\alpha }\equiv dx^{\alpha }/ds$ is the 4-velocity. Furthermore,
suppose the equation of state being of the form $p=\gamma \rho $, then the
solution gives

\begin{eqnarray}
\rho &=&\frac{2}{1+\gamma }\left( \frac{\mu ^{2}+k}{A^{2}}-\frac{\mu \dot{
\mu }}{A\dot{A}}\right) ,  \nonumber \\
\Lambda &=&\frac{2}{1+\gamma }\left[ \left( \frac{1+3\gamma }{2}\right)
\left( \frac{\mu ^{2}+k}{A^{2}}\right) +\frac{\mu \dot{\mu}}{A\dot{A}}\right]
.  \label{origindc}
\end{eqnarray}
We can see that if the two functions $\mu (t)$ and $\nu (t)$ are given, then
the two scale factors $A(t,y)$ and $B(t,y)$ are fixed by (\ref{5Dmetric})
and the mass density $\rho (t,y)$ and the cosmological term $\Lambda (t,y)$
are also fixed by (\ref{origindc}). Generally speaking, on a given $y=const$
hypersurface, $\Lambda =\Lambda (t)$ is a time-variable cosmological term.

From the metric (\ref{5Dmetric}) we see that the form $Bdt=\left( \dot{A}
/\mu \right) dt$ is invariant under an arbitrary coordinate transformation $
t\rightarrow \tilde{t}(t)$. This would enable us to fix one of the two
arbitrary functions $\mu (t)$ and $\nu (t)$, leaving another to account for
the various content of the cosmic matter. Since $\Lambda $\ is a variable
function, one can not simply set $\Lambda =0$ or $\Lambda =$ constant
without losing the generality of the solution. Therefore, to compare the 5D
solution with the $\Lambda =0$ FRW models, a constraint was used in Ref.
[10] that \textit{the cosmological term }$\Lambda $\textit{\ } shown in (\ref%
{origindc}) \textit{\ should decay faster than the density }$\rho $ \textit{%
decreases at late-times of the universe}. Under this constraint, it was
found that the choice $\mu (t)\propto t^{-1/3}$ and $\nu (t)\propto
t^{1/3}$ with $\ k=0$ and $K=1$ gives a matter-dominated model, and the
choice $\mu (t)\propto t^{-1/2}$ and $\nu =const$ with $k=0$ and $K=1$
gives a radiation-dominated model. For these two cases, the universe
evolves, at late-times, with the same rate as in the FRW models.

Here we are going to compare the 5D solution with the present accelerating
universe. We are aware that neither one of the two terms $\Lambda $ and $
\rho $ in (\ref{origindc}) is negligible presently. Therefore we require
that $\Lambda $\textit{\ should decay not faster than }$\rho $\textit{\ does
in the late-times}. In what follows we will show how this is achieved.

Astrophysical data are compatible with a $k=0$ flat 3D space. Meanwhile, we
assume the density $\rho $ being composed mainly of cold dark matter and
baryons. So we choose $k=0$, $K=1$ and $\gamma =0$. Then (\ref{5Dmetric})
and (\ref{origindc}) become 
\begin{eqnarray}
A^{2} &=&(\mu y+\frac{\nu }{\mu })^{2}+\frac{1}{\mu ^{2}}, B=\frac{\dot{A}}{%
\mu },  \nonumber \\
\rho &=&2\left( \frac{\mu ^{2}}{A^{2}}-\frac{\mu \dot{\mu}}{A\dot{A}}\right)
,  \nonumber \\
\Lambda &=&\frac{\mu ^{2}}{A^{2}}+2\frac{\mu \dot{\mu}}{A\dot{A}}.
\label{originabdc}
\end{eqnarray}
Here we explain $\rho $ as the sum of the density parameters of cold dark
matter and baryons, and $\Lambda $ as the dark energy density. We can
designate the corresponding dimensionless densities as $\Omega _{\rho }$ and 
$\Omega _{\Lambda }$, respectively, with $\Omega _{\rho }+\Omega _{\Lambda
}=1$. Then, from (\ref{originabdc}), we obtain

\begin{eqnarray}
\Omega _{\rho } &\equiv &\frac{\rho }{\rho +\Lambda }=\frac{2}{3}\left(1- 
\frac{\dot{\mu}A}{\mu \dot{A}}\right) ,  \nonumber \\
\Omega _{\Lambda} &\equiv &\frac{\Lambda }{\rho +\Lambda }=\frac{1}{ 3}%
\left( 1+\frac{2\dot{\mu}A}{\mu \dot{A}}\right) .  \label{originomegadc}
\end{eqnarray}
Because the 3D space is flat ($k=0$), we assume that both $\mu (t)$ and $\nu
(t)$ can be expressed as power series of $t$. Using the constraint that $
\Lambda $ does not decay faster than $\rho $ does, we consider the following
choice for $\mu (t)$ and $\nu (t)$:

\begin{equation}
\mu (t)=t+\frac{a}{t},\nu (t)=t^{n},  \label{fixedfunctions}
\end{equation}
where $n$ is a parameter to be determined latter. The corresponding form of $
A(t,y)$ is then given by

\begin{equation}
A^{2}=\left[ \left( t+\frac{a}{t}\right) y+\frac{t^{n}}{t+a/t}\right] ^{2}+ 
\frac{1}{\left( t+a/t\right) ^{2}},  \label{exactA}
\end{equation}
If we only consider the condition $n>2$ and the late time of the universe $
t\gg 1$, then (\ref{exactA}) gives

\begin{eqnarray}
A^{2} &=&t^{2n-2}\left[ 1+O\left( t^{-2}\right) \right] ,  \nonumber \\
B &=&\left( n-1\right) t^{n-3}\left[ 1+O\left( t^{-2}\right) \right] .
\label{approximateab}
\end{eqnarray}
Substituting (\ref{fixedfunctions}) and (\ref{approximateab}) into (\ref%
{originabdc}) and (\ref{originomegadc}), we get

\begin{eqnarray}
\rho &=&2\left( \frac{n-2}{n-1}\right) t^{4-2n}\left[ 1+O\left(
t^{-2}\right) \right] ,  \nonumber \\
\Lambda &=&\left( \frac{n+1}{n-1}\right) t^{4-2n}\left[ 1+O\left(
t^{-2}\right) \right] ,  \label{approximatedc}
\end{eqnarray}
and

\begin{equation}
\Omega _{\rho }=\frac{2(n-2)}{3(n-1)}\left[ 1+O\left( t^{-2}\right) \right]
, \Omega _{\Lambda }=\frac{n+1}{3(n-1)}\left[ 1+O\left( t^{-2}\right) \right]
.  \label{approximateomegadc}
\end{equation}
Thus we see that at the late time of the universe $t\gg 1$, $\rho $ and $
\Lambda $ decay with the same rate. This is a very useful result in the
simulation to the present accelerating universe. Remember that we have used
the constraint $n>2$ in (\ref{approximateab}), (\ref{approximatedc}) and (%
\ref{approximateomegadc}). So if $n=2.5$, we have $\Omega _{\rho }\approx
2/9\approx 0.22$ and $\Omega _{\Lambda }\approx 7/9\approx 0.78$. If $n=3$,
we have $\Omega _{\rho }\approx 1/3\approx 0.33$ and $\Omega _{\Lambda
}\approx 2/3\approx 0.67$. Current observation is about $\Omega _{\rho
}\approx 0.3$ and $\Omega _{\Lambda }\approx 0.7$. So the parameter $n$ lies
between $2.5$ and $3$. For an illustration we let $n=3$. Then (\ref{exactA})
gives an exact solution being

\begin{equation}
A^{2}=\left[ \left( t+\frac{a}{t}\right) y+\frac{t^{3}}{t+a/t}\right] ^{2}+ 
\frac{1}{\left( t+a/t\right) ^{2}}, B=\frac{\dot{A}}{\mu }.  \label{exactAB}
\end{equation}
For $t\gg 1$, this exact solution gives

\begin{eqnarray}
A &=&t^{2}\left[ 1+O\left( t^{-2}\right) \right] ,B=\frac{ \dot{A}}{\mu }=2%
\left[ 1+O\left( t^{-2}\right) \right] ,  \nonumber \\
\rho &=&t^{-2}\left[ 1+O\left( t^{-2}\right) \right] ,\Lambda =2t^{-2}\left[
1+O\left( t^{-2}\right) \right] ,  \nonumber \\
\Omega _{\rho } &=&\frac{1}{3}\left[ 1+O\left( t^{-2}\right) \right] ,
\Omega _{\Lambda }=\frac{2}{3}\left[ 1+O\left( t^{-2}\right) \right] .
\label{allapproximate}
\end{eqnarray}
Figure 1 is a plot of the scale factor $A\left( t,y\right) $ of the exact
solution (\ref{exactAB}) on the $y=1$ hypersurface (with $a=1$). From this
figure we can see that there is a finite minimum for the scale factor $
A\left( t\right) $ at $t=t_{b}$ which represents a ``big bounce''. Before it
the universe contracts, and after it the universe expands.

\begin{figure}[tbp]
\centering\includegraphics[width=2.5in,height=2.5in]{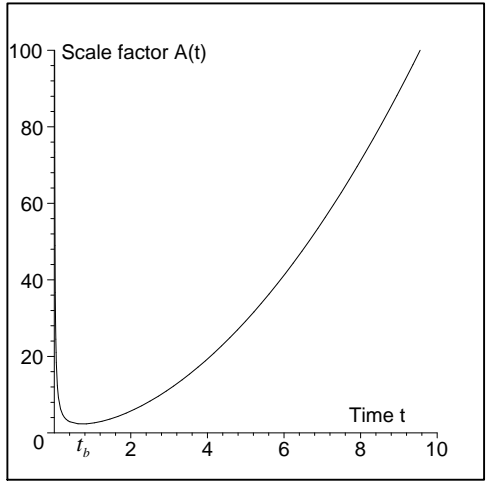}
\caption{Evolution of the scale factor $A(t)$ with $
A^{2}=[(t+1/t)+t^{3}(t+1/t)^{-1}]^{2}+(t+1/t)^{-2}.$ There is a minimum at $
t=t_{b}$ which corresponds to a bounce.}
\end{figure}

To see clearly how the universe evolves, we need an appropriate definition
for the Hubble and the deceleration parameters. Consider the metric (\ref%
{5Dmetric}). On a given $y=const$ hypersurface the proper time can be
defined as $d\tau =Bdt$. So the invariant definitions for the Hubble and
deceleration parameters should be given as$^{21}$ 
\begin{eqnarray}
H(t,y) &\equiv &\frac{1}{A}\frac{dA}{d\tau }=\frac{1}{B}\frac{\dot{A}}{A}=%
\frac{\mu }{A},  \nonumber \\
q(t,y) &\equiv &\left. -A\frac{d^{2}A}{d\tau ^{2}}\right/ \left( \frac{dA}{%
d\tau }\right) ^{2}=-\frac{A\dot{\mu}}{\mu \dot{A}}.  \label{exactHq}
\end{eqnarray}%
Then, with use of (\ref{fixedfunctions}) and (\ref{allapproximate}),
equation (\ref{exactHq}) yields, for $t\gg 1$,

\begin{eqnarray}
H &=&\frac{1}{t}\left[ 1+O\left( t^{-2}\right) \right] ,  \nonumber \\
q &=&-\frac{1}{2}\left[ 1+O\left( t^{-2}\right) \right] .
\label{approximateHq}
\end{eqnarray}
Thus we find that at the late time of the universe $t\gg 1$, the universe 
(\ref{exactAB}) is accelerating with $q\approx -1/2.$

Using (\ref{allapproximate}) in the definition of the proper time $d\tau
=Bdt $, we find

\begin{equation}
\tau -\tau _{b}\approx 2t  \label{ttall}
\end{equation}
for $t\gg 1$ and $\tau \gg \tau _{b}$, where $\tau _{b}$ represents the
initial bounce time. So the five-dimensional line element (\ref{5Dmetric})
gives

\begin{equation}
dS^{2}\longrightarrow d\tau ^{2}-\frac{1}{16}\left( \tau -\tau _{b}\right)
^{4}\left( dr^{2}+r^{2}d\Omega ^{2}\right) -dy^{2}  \label{approximatemetic}
\end{equation}
for $\tau \gg \tau _{b}$. This is an approximate metric to describe our
present accelerating universe.

\begin{figure}[tbp]
\centering \includegraphics[width=2.5in,height=2.5in]{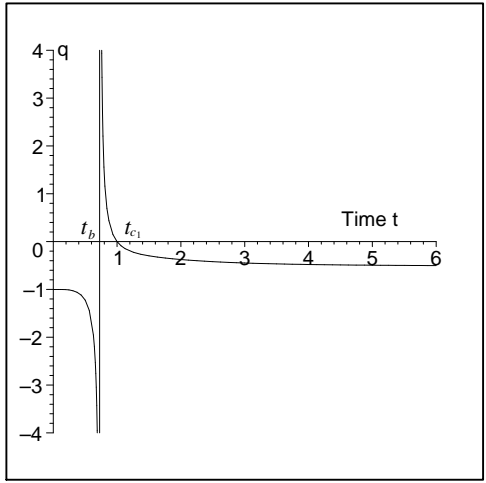}
\caption{Global evolution of the deceleration parameter $q(t,y)$ of the
solution (12) with $a=1$ and $y=1$. The bouncing time is at $t=t_{b}$. There
is a critical time $t_{c1\text{ }}$ before which the universe is
decelerating and after which the universe is accelerating.}
\end{figure}

\begin{figure}[tbp]
\centering\includegraphics[width=2.5in,height=2.5in]{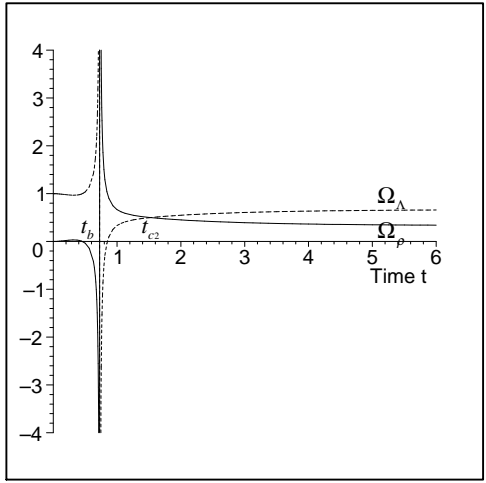}
\caption{Global evolutions of $\Omega _{\protect\rho }$ and $\Omega
_{\Lambda }$ of the solution (12) with $a=1$ and $y=1$. The solid line
represents $\Omega _{\protect\rho }$ and the dashed line represents $\Omega
_{\Lambda }$. There is a critical time $t_{c2}$ at which $\Omega _{\Lambda }$
takes over $\Omega _{\protect\rho }$ and dominates the universe.}
\end{figure}

For the special exact solution (\ref{exactAB}), we can obtain an exact
expression for the deceleration parameter $q(t,y)$ on a given $y=const$
hypersurface by substituting (\ref{fixedfunctions}) and (\ref{exactAB}) in
the second equation of (\ref{exactHq}). This is a long expression and we
plot it in Figure 2 with $a=1$ and $y=1$.

Similarly, we can use the exact solutions (\ref{exactAB}) and (\ref%
{fixedfunctions}) in the definition (\ref{originomegadc}) to obtain exact
expressions for $\Omega _{\rho }$ and $\Omega _{\Lambda }$. The global
evolutions of $\Omega _{\rho }$ and $\Omega _{\Lambda }$ are ploted in
Figure 3 with $a=1$ and $y=1$.

From Fig. 2 and Fig. 3 we can see clearly that the global evolution of the
universe is divided into four stages separated by the ``big bounce'' point $%
t_{b}$ and two critical points $t_{c1}$ and $t_{c2}$. The first stage is at $%
0<t<t_{b}$ in which the universe contracts from a $\Lambda $ -dominated
vacuum. The bounce point $t_{b}$ is a matter singularity (see also Refs.
[10,15,21]), across\textbf{\ }which $\Omega _{\rho }$ jumps from $-\infty $
to $+\infty $ and $\Omega _{\Lambda }$ jumps from $+\infty $ to $-\infty $.
The second stage is at $t_{b}<t<t_{c1}$ in which $\Omega _{\rho }$ decreases
and $\Omega _{\Lambda }$ increases with $\Omega _{\rho }>\Omega _{\Lambda }$
and $q>0$, and the universe is decelerating. The third stage is at $%
t_{c1}<t<t_{c2}$ in which the deceleration parameter $q$ becomes negative,
implying that the expansion of the universe turns to speed up. The fourth
stage is at $t_{c2}<t$ in which $q<0$, so the universe is accelerating. In
this stage, $\Omega _{\Lambda }$ overtakes $\Omega _{\rho }$ and the
universe is dominated by dark energy. For $t\gg 1$ we have $\Omega _{\Lambda
}/\Omega _{\rho }\longrightarrow 2,$ in agreement approximately with
observations of the present stage of our universe.

\section{Discussion}

The ``big bounce'' solution (\ref{5Dmetric}) is characterized by having a
``bounce'', rather than a ``bang'', as the ``beginning'' of our expanding
universe, and by having a evolving cosmological ``constant''.
Mathematically, the general solution (\ref{5Dmetric}) contains two arbitrary
functions $\mu (t)$ and $\nu (t)$. Different choices of these two
functions may give different models to describe different stages of our
universe. In this paper we find that a simple choice of them yields a simple
exact solution (\ref{exactAB}) which can describe the present accelerating
universe in a satisfactory manner.

We should emphasize that although this simple exact solution exhibits many
interesting features of the bounce model, it has to be generalized to meet
more observations and to explain the bouncing from a physical point of view.
For instance, the two arbitrary functions $\mu (t)$ and $\nu (t)$ might
be not as simple as given in (\ref{fixedfunctions}) for a real cosmological
model. Meanwhile, a variable cosmological ``constant'' might be also too simple 
to describe dark energy. It is known from literatures$^{8,9,22-24}$ that a properly 
chosen scalar field might be more suitable to describe dark energy. Be aware that 
our model (\ref{5Dmetric}) is 5D empty and 4D sourceful with an induced 4D energy-momentum 
tensor which has been supposed to contain a cosmological term $\Lambda $ as shown in (\ref{Einsteintensor:}). 
Thus, instead of a $\Lambda $ term, we probably should consider the case where the induced matter contains a scalar field as 
a component. We wish this scalar field may provide us with a mechanics to explain the acceleration 
as well as the bouncing of the universe. We are going to do this in future studies.

We thank Baorong Chang, Huanying Liu and Feng Luo for discussion. This work was
supported by NSF of P. R. China under Grant 10273004.

\end{document}